\documentclass[11pt]{article}
\usepackage{epsfig}
\usepackage{epstopdf}
\usepackage{amsmath}\usepackage{amssymb}
\usepackage{mathrsfs}
\usepackage{graphicx}
\usepackage{float}
\usepackage{subcaption}
\usepackage{caption}

\newcommand {\be}{\begin{equation}}
\newcommand {\ee}{\end{equation}}
\newcommand {\ba}{\begin{eqnarray}}
\newcommand {\ea}{\end{eqnarray}}

\begin{document}
	\def \a'{\alpha'}
	\baselineskip 0.65 cm
	\begin{flushright}
		\ \today
	\end{flushright}

\begin{center}{\large
		{\bf Searching for Vector-Like Quarks in a Fermionic Dark Matter Model with Pseudoscalar: A Resonance Case}} {\vskip 0.5 cm} {\bf ${\rm Seyed~ Yaser~ Ayazi}$$^1$, ${\rm  Ahmad~ Mohamadnejad}$$^2$ and ${\rm S.~Peyman~ Zakeri}$$^{3,4}$}{\vskip 0.5 cm}
	{\small $^1$$Faculty~of~Physics$,~		$Semnan~University$~$P.O.~Box~35195-363$,~$Semnan$,~$Iran$
		$^2$$Young~Researchers~and~Elite~Club$,~$Islamshahr~Branch$,~$Islamic~Azad~University$,
		~$Islamshahr~3314767653$,~$Iran$
		$^3$$Faculty~of~Physics$,~ $Yazd~University$,~$P.O.~Box~89195-741$,~$Yazd$,~$Iran$
			$^4$$School~ of~ Particles~ and~ Accelerators,~ Institute~ for~ Research~ in~
			Fundamental$ $ Sciences~(IPM), P.O.~Box~ 19395-5531, Tehran, Iran$}
\end{center}

\begin{abstract}
We propose a minimal model of a fermionic dark matter with a pseudoscalar mediator and N generation of vector-like quarks. We calculate the relic density and obtain new constraint on the generation of the aforementioned quarks. Concerning phenomenological aspects, we probe the presenting model via direct and indirect approaches. Finally, as an illustrative example, we evaluate a resonance case which has been (would be) the subject of major experiments aiming to detect new particles. Performing this analysis results in significant constraints on the coupling strength of the vector-like quarks.       
\end{abstract}

\section{Introduction} \label{sec1}
A body of evidence claims that our Universe is filled with a mysterious Dark Matter (DM), which may be a new kind of particle embedded beyond the Standard Model (SM). Weakly Interacting Massive Particles (WIMPs) \cite{Gondolo:1990dk,Srednickii,Chiu,Ayazi:2015mva} are the most popular solution of DM puzzle. Apart from that, Feebly Interacting Massive Particles (FIMPs) \cite{McDonald:2001vt,Hall:2009bx,Ayazi:2015jij,Bernal:2017kxu,Zakeri:2018hhe} make the only reliable framework to explain this non-visible particle. Beside the efforts to detect the nature of DM, it has been proved that its features can help us to explain new particles beyond the SM. Vector-like lepton (VLL) \cite{Nakamura:2010zzi,Aad:2015dha} and vector-like quark (VLQ)  \cite{Ellis:2014dza,Aguilar-Saavedra:2013qpa,Azatov:2012rj,Angelescu:2015uiz} are two relevant examples which the latter one is the subject of current paper.

In the extension of the SM, VLQs are the heavy colored particles which are triplets under the color gauge group and have vector-like couplings. The left-handed and right-handed chiralities of VLQs transform in the same way under the SM gauge groups ${\rm SU(3)}_C \times {\rm SU(2)}_L \times {\rm U(1)}_Y$. They can mix with the SM quarks and have both left-handed and right-handed charged currents. Study of VLQs, is well-motivated due to the following reasons: explaining new sources for CP violation \cite{delAguila:1997vn}, emerging as fermion resonances \cite{Contino:2006nn} and phenomenological investigation of pseudo-Goldstone Higgs boson \cite{Perelstein:2003wd}.            

In this paper, we investigate whether a fermionic DM in the frame of WIMP can be used to explain VLQs. In this way, a Dirac particle (DM) interacts with the visible sector mediated by a pseudoscalar. This mediator is related to N generation of VLQs by Yukawa couplings. We will show that our model can obtain the correct relic density of DM together with significant constraints on VLQs generations and couplings. This has been complimentarily done by use of a resonance example.    
  
Recording resonances in the diphoton events is one the most important goals followed at the LHC. To this aim, LHC Runs 1, 2 has performed in the past years \cite{Aad:2012tfa, Chatrchyan:2012xdj, ATLAS, CMS}; The outputs was respectively the valuable discovery of the Higgs boson in 2012 \cite{Aad:2012tfa, Chatrchyan:2012xdj} and a controversial excess in 2015 \cite{ATLAS, CMS}. Although the latter was washed out by ATLAS 2016 report, searching for new particles still continues in this way. Thus, it is worth nothing here that we consider the resonance case of such fermionic model for future applications in next particle physics.

This paper is organized as follows: in section 2, we introduce our model in which a Dirac fermion plays the role of DM and interacts with the SM through the Higgs portal. In section 3, we obtain the relic density of DM and study its direct and indirect detections in the following two sections. Then, we analyze a resonance case and probe the parameter space of the VLQs more exhaustively. Finally, we present our results in section 7.

\section{The Model} \label{sec2}
In this section, we construct our model by considering an extension of the SM through three new fields ($\chi, \phi$ and $\psi$). In this setup, we introduce N generation of VLQs, denoted by $\psi$, and a Dirac spinor $\chi$ which plays the role of DM.
VLQs are colored and carry the electro-magnetic charge and $\phi$ and $\chi$ are singlet under the SM gauge groups. We also suppose that $\chi$ is odd under a $Z_{2}$ symmetry which guarantees its stability as a DM candidate \textendash all other SM fields are even under this discrete symmetry. In addition, to avoid any mixing between VLQs and their SM counterparts, we consider them odd under the $Z_2$ symmetry. The transformations of the new fields have been summarized in Table.~\ref{fields}.

\begin{table*}
	\begin{center}
		\begin{tabular}{c|c|c|c|c}
			Fields & ${\rm SU(3)}_C$ & $ {\rm SU(2)}_L$ & ${\rm U(1)}_Y$ & $Z_2$ \\
	     \hline
	     	$\phi$ & 1  & 1 & 0 & + \\
	       $\chi$ & 1 & 1 & 0 & - \\
		   $\psi$ & 3 & 1 & 2/3 & - \\
		\end{tabular}
	\end{center}
	\caption{Transformations of the new fields under the SM gauge groups and the $Z_2$ symmetry.}
	\label{fields}
\end{table*}

Considering the framework introduced above, the potential for the new fields which is renormalizable and invariant under the gauge and $ Z_{2} $-symmetry is given by:  
\begin{align} \label{lagrangian}
V(\phi,\psi,\chi)&=-ig_{\phi \chi} \phi \overline{\chi} \gamma^{5} \chi
 + ig_{\phi \psi} \phi \overline{\psi} \gamma^{5} \psi
 -Q \overline{\psi}\gamma^{\mu}\psi A_{\mu}
 + g_{s} \overline{\psi} \lambda^{a} \gamma^{\mu}\psi G_{\mu}^{a} \nonumber \\
&- \frac{1}{2} \mu_{\phi}^{2} \phi^{2} - \frac{1}{4!} \lambda_{\phi} \phi^{4}
 - \lambda_{\phi H} \phi^{2}  H^{\dagger}H\ .
 \end{align}
 where $ Q=\frac{2}{3}$ or $-\frac{1}{3} $ is the electromagnetic charge,
 $ g_{s} $ is the strong coupling constant and $\lambda^{a} (a=1, 2, 3, . . . , 8)$ denotes the Gell-Mann matrices. As we demand ${\cal L}$ to be CP-invariant, the Lagrangian does not include $\phi$, $\phi^3$ and $\phi H^2$ terms. Therefore, our model has the minimal interactions in comparison with the models featuring scalar mediator in which the aforementioned interaction terms appear. In principle, $\phi$ can acquire a Vacuum Expectation Value (VEV), but demanding our model to be invariant under CP after Spontaneous Symmetry Breaking (SSB), we take the VEV of the pseudoscalar equal to zero. Thus we expect that after SSB, the mass matrix of the fields $\phi$ and $H$ to be diagonal.

Here, in addition of the SM parameters, we consider 5 new ones as $M_\phi$, $M_{\chi}$, $M_{\psi}$ (mass parameters), $g_{\phi\psi}$ and $g_{\phi\chi}$ (coupling parameters) which are independent free parameters of the model. In what follows, we will probe them experimentally in order to reach the desired parameter space.

\section{Relic Abundance} \label{sec3}
To calculate the relic density of DM, we have applied the relevant version of micrOMEGAs package \cite{micromegas}. This has been experimentally done by Planck measurement \cite{PLANCK} and the current value is $\Omega_{\rm DM}h^2 = 0.1199 \pm 0.0027$
where $h = 0.67 \pm 0.012$ is the scaled current Hubble parameter in units of $100~\rm km/s.Mpc$.
The current relic density of DM can be obtained from:
\begin{equation}
\Omega_{\rm DM}h^2 =2.742\times10^{-8}(\frac{M_{\chi}}{\rm GeV}) Y(T_0).
\label{Re}
\end{equation}

\begin{figure}
	\begin{center}
		\centerline{\hspace{0cm}\epsfig{figure=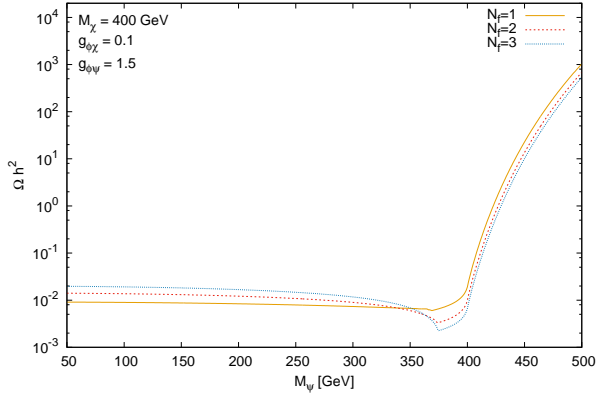,width=8cm}}
		\centerline{\vspace{-0.8cm}}
	\end{center}
	\caption{Relic density of DM as a function of VLQs mass for different generations of VLQs.}\label{relic}
\end{figure}

Fig.~\ref{relic} indicates how our model contributes to the DM density of the Universe for different generations of VLQs and proper values of the couplings. Here, we have chosen the mediator mass as $M_\phi=750$ GeV which is useful for further application in section~\ref{sec6}. From the figure, it can be seen that for VLQs mass larger than $M_{\phi}/2$, the contribution to the relic density will be increased. This is  due to the production of DM through the probable channel $\psi\psi\rightarrow \phi\rightarrow\chi\chi$. 

\begin{figure}
	\centering
	\begin{subfigure}[t]{0.53\textwidth}
		\includegraphics[width=1\linewidth, height=5.0cm]{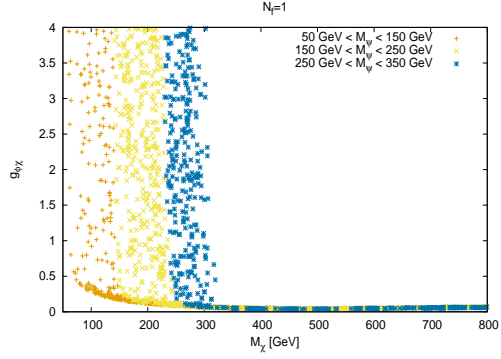} 
		\caption{}
	\end{subfigure}
    \hfill
	\begin{subfigure}[t]{0.53\textwidth}
		\raisebox{-\height}{\includegraphics[width=1\linewidth, height=5.0cm]{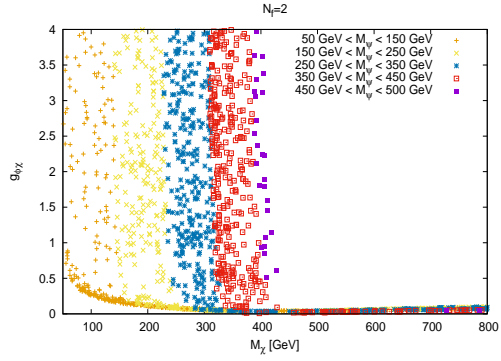}}
		\caption{}
	\end{subfigure}
	\begin{subfigure}[t]{0.53\textwidth}
	\raisebox{-\height}{\includegraphics[width=1\linewidth, height=5.0cm]{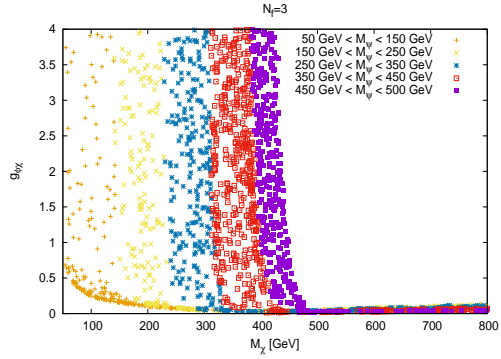}}
	\caption{}
	\end{subfigure}
	\caption{Scater points depict ranges of parameter space in the ($M_\chi, |g_{\phi\chi}|$)-plane which are consistent with the observed value of the DM density. In this figure, we set different intervals of VLQs mass and take its generation as a) $\rm N_f=1$, b) $\rm N_f=2$ and c)$\rm N_f=3$.}
	\label{scater-relic}
\end{figure}

In Fig.~\ref{scater-relic}, we have shown allowed regions in the plane of DM mass and its coupling to the mediator which are consistent with the observed relic density. In this analysis, we have considered VLQ generations ($\rm N_f$) varying from 1 to 3 and have chosen various intervals of VLQs  mass. It could be derived from the figure that large values of DM mass makes the parameter space larger although employs small values of coupling $g_{\phi\chi}$. But, as the generation, $\rm N_f$, increases from 1-3, we can construct a proper space including larger values of $g_{\phi\chi}$. Therefore, reaching to the observed relic density, indicates the role of VLQ generation.

\section{Direct Detection} \label{sec4}
In our model, we have considered a singlet fermionic field as a DM candidate. Since the pseudo scalar does not mix with the SM Higgs,  DM can not interact with nucleon by Higgs boson exchange. Therefore, there is no detectable signal in the elastic scattering of DM off nuclei. Subsequently, our model evades from direct detection constraints coming from experiments such as XENON100 \cite{Aprile:2012nq} and LUX \cite{Akerib:2013tjd}.

\section{Indirect Detection} \label{sec5}
\begin{figure}
	\begin{center}
		\centerline{\hspace{1cm}\epsfig{figure=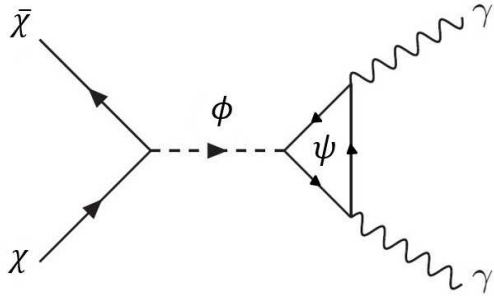,width=7cm}}
		\centerline{\vspace{-0.8cm}}
	\end{center}
	\caption{The dominant Feynman diagram for DM annihilation into monochromatic gamma ray lines.}\label{indirect}
\end{figure}

Concerning indirect detection, we calculate the possible contribution to the annihilation of DM to $\gamma\gamma$. Fermi-LAT collaboration \cite{Fermi-Lat} has reported an upper limit for the thermal average of cross section for this process. For DM mass of ${\cal O} (100)~ \rm GeV$, this experimental report lies in the range $10^{-29}-10^{-27}~\text{cm}^3\text{s}^{-1}$. As it is shown in Fig. \ref{indirect}, the aforementioned process is mediated by the pseudoscalar $\phi$ and the vertex $\phi\gamma\gamma$ can be generated at the quantum level with a loop involving the massive vector-like field $\psi$. Therefore, we can obtain the following expression for the thermal averaged cross section:
\begin{eqnarray}
\langle\sigma v\rangle_{\gamma\gamma}=\frac{\alpha_e^2 g_s^2s^2g_{\phi\chi}^2g_{\phi\psi}^2}{64\pi^3M_W^2M_\psi^2}\frac{1}{(s-M_\phi^2)^2+M_\phi^2\Gamma_\phi^2}\arrowvert A_{1/2}(\tau_\psi)\arrowvert^2,
\end{eqnarray}
where $\alpha_e$ is the electromagnetic coupling constant, $g_s$ the strong one and $\sqrt{s}$ is the center of mass energy. $A_{1/2}(\tau_\psi)$ denotes the form factor and is given by   
\begin{eqnarray} \label{eq:4}
A_{1/2}(\tau_\psi)=-2\tau(1+(1-\tau)f(\tau)),
\end{eqnarray}
where
\begin{eqnarray} \label{Eq:7}
f(\tau) =
\left\{
\begin{array}{ll}
(\sin^{-1}\sqrt{1/\tau})^2  & \mbox{if } \tau \geq 1, \\
-\frac{1}{4}(\log\frac{1+\sqrt{1-\tau}}{1-\sqrt{1-\tau}}-i\pi)^2 & \mbox{if } \tau < 1.
\end{array}
\right.
\end{eqnarray}
For the case of presenting model, the limit employing large values of the couplings cannot exceed $10^{-31}~\text{cm}^3\text{s}^{-1}$ which is very smaller than that of Fermi-LAT. Anyway, the sommerfeld enhancement \cite{Feng:2010zp,Finkbeiner:2010sm,Lu:2017jrh} may increase DM annihilation cross section in order to reach the experimental upper limits detected by Fermi-LAT.              

\section{Example: Diphoton Resonance} \label{sec6}
In this section, we consider a resonance case inspired by the events that has occurred at the LHC. This probe may be useful to explain resonance data in the future. As a practical example, we consider the diphoton excess reported in 2015; the ATLAS and CMS collaborations reported excesses with a mass of about 750 $\rm GeV$ and a decay width of 45 $\rm GeV$ \cite{ATLAS, CMS} at proton center of mass energy $13~\rm TeV$.

Following the procedure of these experiment, we analyze the diphoton production by interpreting the observed excess as a resonance in the process $pp\rightarrow\phi\rightarrow\gamma\gamma$. Here, $\phi$ coincides with the pseudoscalar presented in the context of our model. The ATLAS and CMS collaborations have observed the diphoton excess respectively with $3.6~ \sigma$ and $2.6~\sigma$ deviation from the central value of the SM prediction at $750~\rm GeV$. This means that $\sigma(pp\rightarrow\phi\rightarrow\gamma\gamma)$ is found in the order of $5-10~ \rm fb$ which is consistent with the combination of both experiments. In this production cross section, the gluon fusion is dominant and we can describe the aforementioned process by the decays $\phi\rightarrow gg$ and  $\phi\rightarrow \gamma\gamma$ at $\sqrt{s}=13~\rm GeV$:
\begin{eqnarray}
\sigma(pp\rightarrow\phi\rightarrow \gamma\gamma)&=&\sigma(pp\rightarrow\phi)\text{Br}(\phi\rightarrow \gamma\gamma)\nonumber \\
&=&\frac{C_{gg}}{sM_{\phi}\Gamma_{tot}}\Gamma(\phi\rightarrow gg)\Gamma(\phi\rightarrow \gamma\gamma),
\end{eqnarray}
where $\Gamma_{tot}$ is the total decay width of the pseudoscalar field and 
$C_{gg}$ is the parton distribution function (PDF) of the gluon:
\begin{eqnarray}
C_{gg}=\frac{\pi^2}{8}\int_{M_\phi^2/s}^{1}\frac{dx}{x}g(x)g(\frac{M_\phi^2}{sx}).
\end{eqnarray}
The partial decay widths of $\phi$ are given by:
\begin{eqnarray}
\Gamma(\phi\rightarrow gg)&=&\frac{\alpha_s^2 M_{\phi}^3 }{128\pi^3}|\frac{g_{\phi\psi}}{M_{\psi}}A_{1/2}(\tau_{\psi})|^2,\\
\Gamma(\phi\rightarrow \gamma\gamma)&=&\frac{\alpha_e^2 M_{\phi}^3 }{1024\pi^3}|2N_cQ^2\frac{g_{\phi\psi}}{M_{\psi}}A_{1/2}(\tau_{\psi})|^2,
\end{eqnarray}
where $\tau_\psi=\frac{4M_\psi^2}{M_\phi^2}$ and $A_{1/2}(\tau_\psi)$ was stated in Eq. \ref{eq:4}.

Large value of the decay width in the dipoton excess reported by ATLAS collaboration, invites us to consider the possibility of decaying the pseudoscalar into the new particles. The total decay width of $\phi$ is then given by:
\begin{eqnarray}
\Gamma_{tot}=\Gamma(\phi\rightarrow\psi\psi)+\Gamma(\phi\rightarrow\chi\chi)+\Gamma(\phi\rightarrow gg)+\Gamma(\phi\rightarrow \gamma\gamma),
\end{eqnarray}
where $\Gamma(\phi\rightarrow \chi\chi)$ and $\Gamma(\phi\rightarrow \psi\psi)$ could be written by:
\begin{eqnarray}
\Gamma(\phi\rightarrow \chi\chi)=\frac{M_{\phi} g_{\phi\chi}^2}{8\pi} (1-4M_\chi^2/M_{\phi}^2)^\frac{1}{2},
\end{eqnarray}
\begin{eqnarray}
\Gamma(\phi\rightarrow \psi\psi)=\frac{M_{\phi}g_{\phi\psi}^2}{8\pi }(1-4M_{\psi}^2/M_{\phi}^2)^\frac{1}{2}.
\end{eqnarray}
It should be noted here that to explain the decay width of the diphoton excess in the order of $10 ~\rm GeV$, we have to consider invisible decay mode $\phi\rightarrow \chi\chi$; otherwise the processes $\phi\rightarrow \gamma\gamma$ and $\phi\rightarrow gg$ can not satisfy the ATLAS value of the decay width.

\begin{figure}
	\centering
	\begin{subfigure}[t]{0.53\textwidth}
		\includegraphics[width=1\linewidth, height=5.0cm]{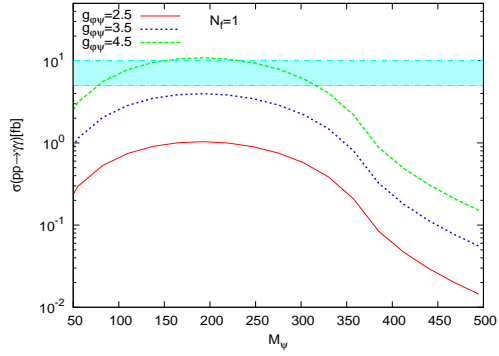} 
		\caption{}
		\end{subfigure}
	\hfill
	\begin{subfigure}[t]{0.53\textwidth}
		\raisebox{-\height}{\includegraphics[width=1\linewidth, height=5.0cm]{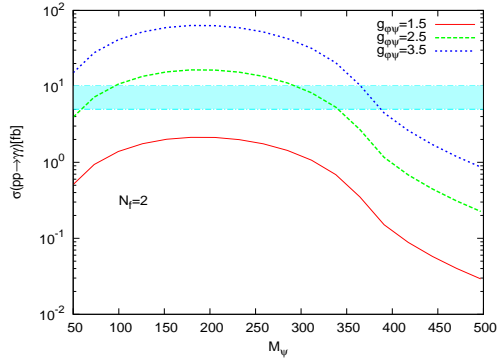}}
		\caption{}
		\end{subfigure}
    \begin{subfigure}[t]{0.53\textwidth}
	   \raisebox{-\height}{\includegraphics[width=1\linewidth, height=5.0cm]{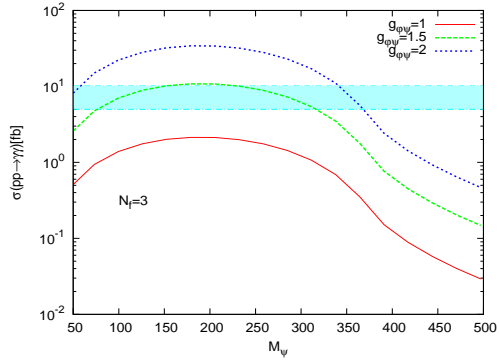}}
	   \caption{}
	\end{subfigure}
	\caption{$\sigma(pp\rightarrow \gamma\gamma)$ as a function of VLQ mass for different values of $g_{\phi\psi}$ regarding possible generations of VLQs.}
	\label{cross}
\end{figure}

In Fig.~\ref{cross}, we displayed the production cross section of the diphoton resonance as a function of VLQs mass for different values of the coupling $g_{\phi\psi}$ and different numbers of VLQs generation. Cyan strip shows the bound of the cross section at the LHC between $5$ to $10~\rm fb$.  
For the sake of simplicity, we have supposed here that all generations of VLQs have the same mass and coupling with the pseudoscalar $\phi$. As is seen in Fig.~\ref{cross}, for more number of VLQ generations, the LHC bound is saturated by smaller values of coupling $g_{\phi\psi}$.

\begin{figure}
	\centering
	\begin{subfigure}[t]{0.53\textwidth}
		\includegraphics[width=1\linewidth, height=5.0cm]{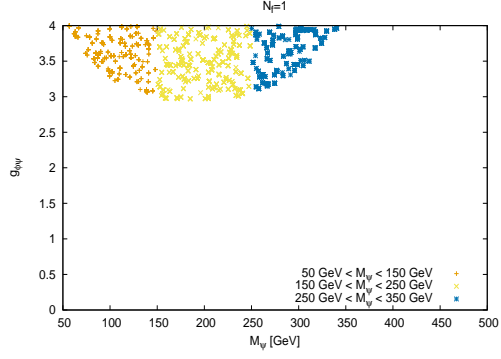} 
		\caption{}
	\end{subfigure}
	\hfill
	\begin{subfigure}[t]{0.53\textwidth}
		\raisebox{-\height}{\includegraphics[width=1\linewidth, height=5.0cm]{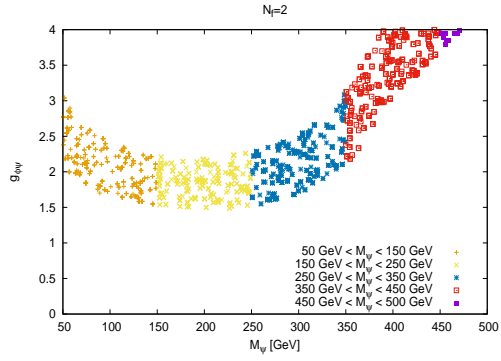}}
		\caption{}
	\end{subfigure}
	\begin{subfigure}[t]{0.53\textwidth}
		\raisebox{-\height}{\includegraphics[width=1\linewidth, height=5.0cm]{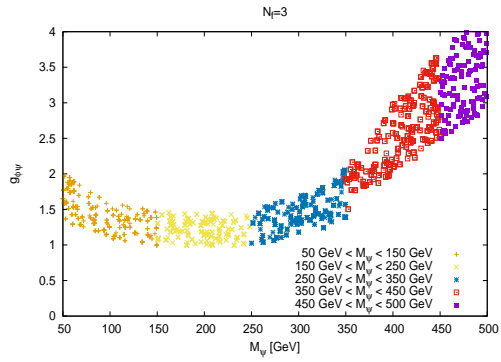}}
		\caption{}
	\end{subfigure}
	\caption{Scater points depict ranges of parameter space in ($M_\psi, |g_{\phi\psi}|$)-plane which are consistent with the cross section of the diphoton resonance for 1 to 3 generations of VLQs.}
	\label{scatercross}
\end{figure}

\begin{figure}
	\begin{center}
		\centerline{\hspace{-0.5cm}\epsfig{figure=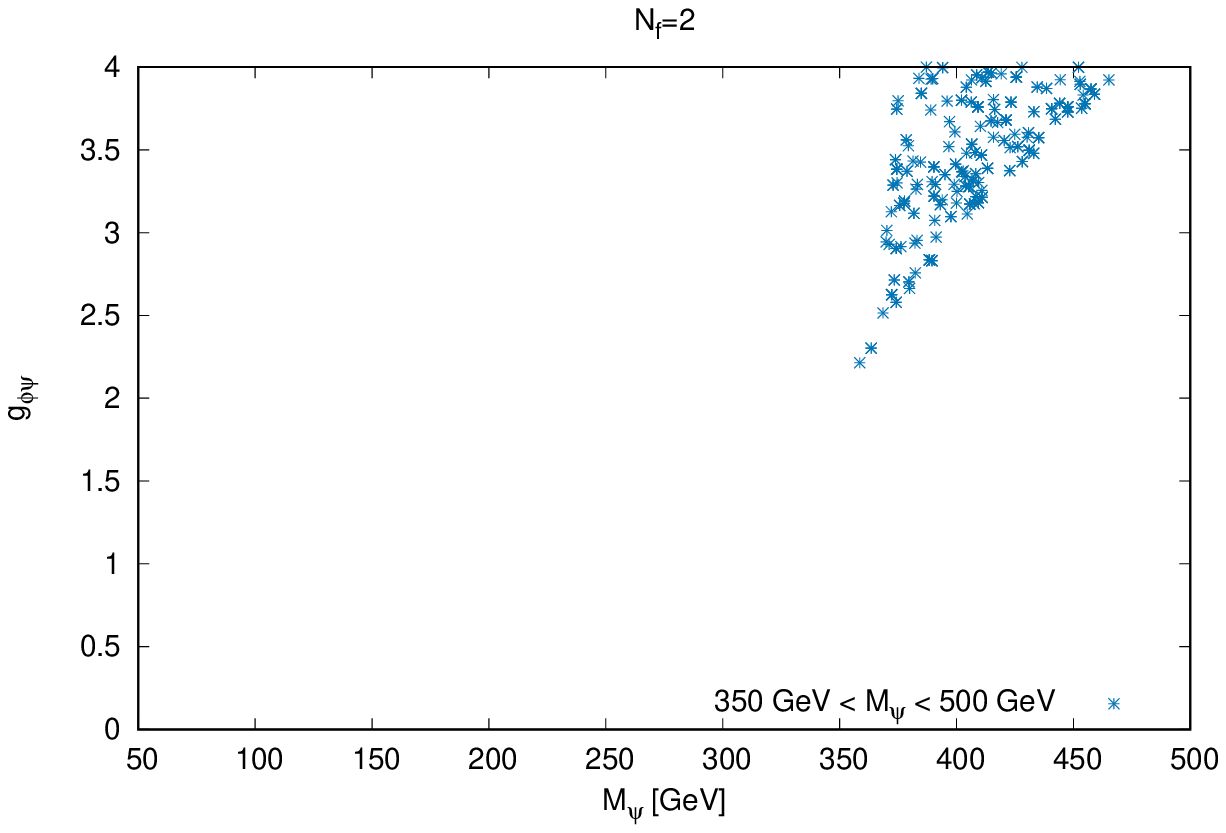,width=6.5cm}\hspace{0cm}\epsfig{figure=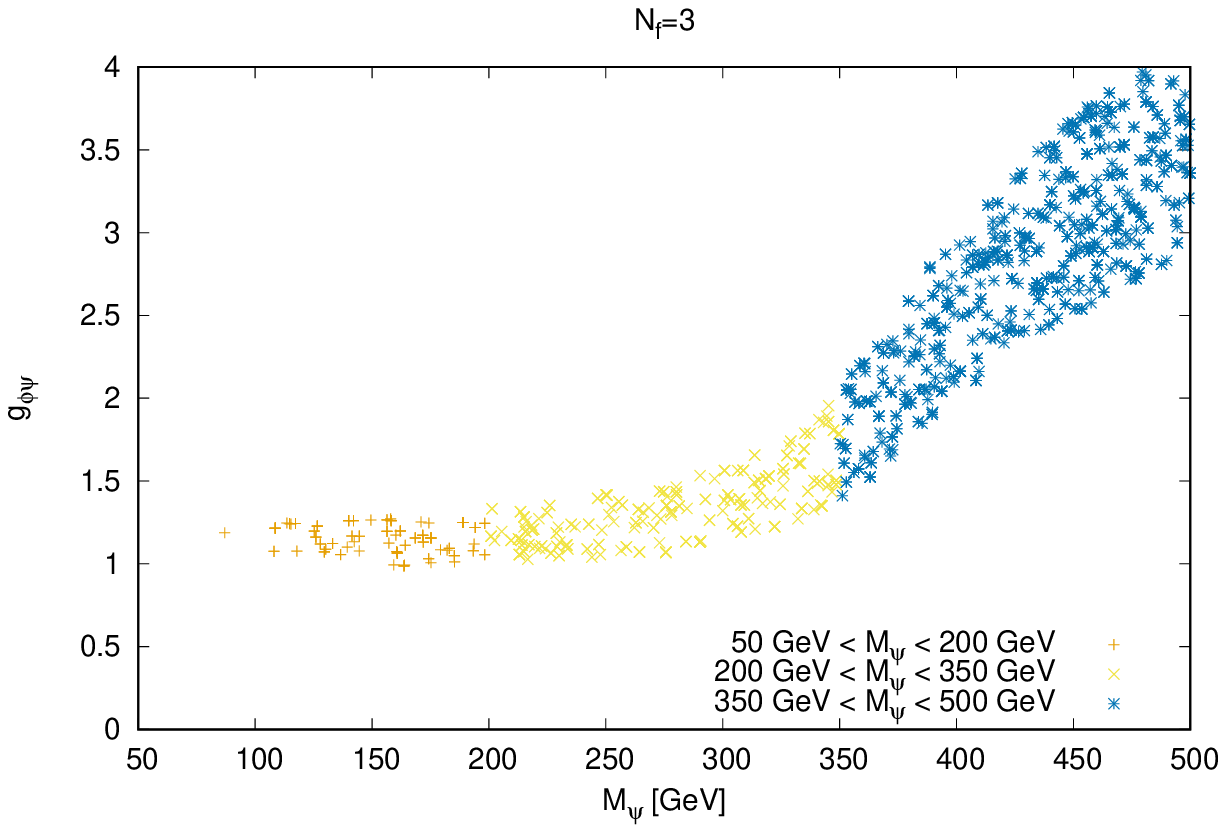,width=6.5cm}}
		\centerline{\vspace{-1.5cm}\hspace{0.5cm}(a)\hspace{6cm}(b)}
		\centerline{\vspace{0.5cm}}
	\end{center}
	\caption{Similar to Fig.~\ref{scatercross} except, we also consider the condition of the diphoton excess width at the ATLAS.}\label{scater-crosswidth}
\end{figure}

Fig.~\ref{scatercross} shows allowed scattered areas of the parameter space in $M_\psi$ and $|g_{\phi\psi}|$ plane which are consistent with the cross section of the diphoton resonance for 1-3 generations of VLQs. We have indicated VLQs mass in the different colors for a better comparison with the last analysis which has been done in the previous section. In Fig.~\ref{scater-crosswidth}, we also considered the condition of the diphoton excess width recorded at the ATLAS. As it can be seen, allowed region has shrunk significantly in this case and the first generation of VLQs doesn't contribute anymore. In addition, increasing the VLQs mass, we reach to a larger parameter space which accompanies with large coupling constant. 

\begin{figure}
	\begin{center}
		\centerline{\hspace{-0.5cm}\epsfig{figure=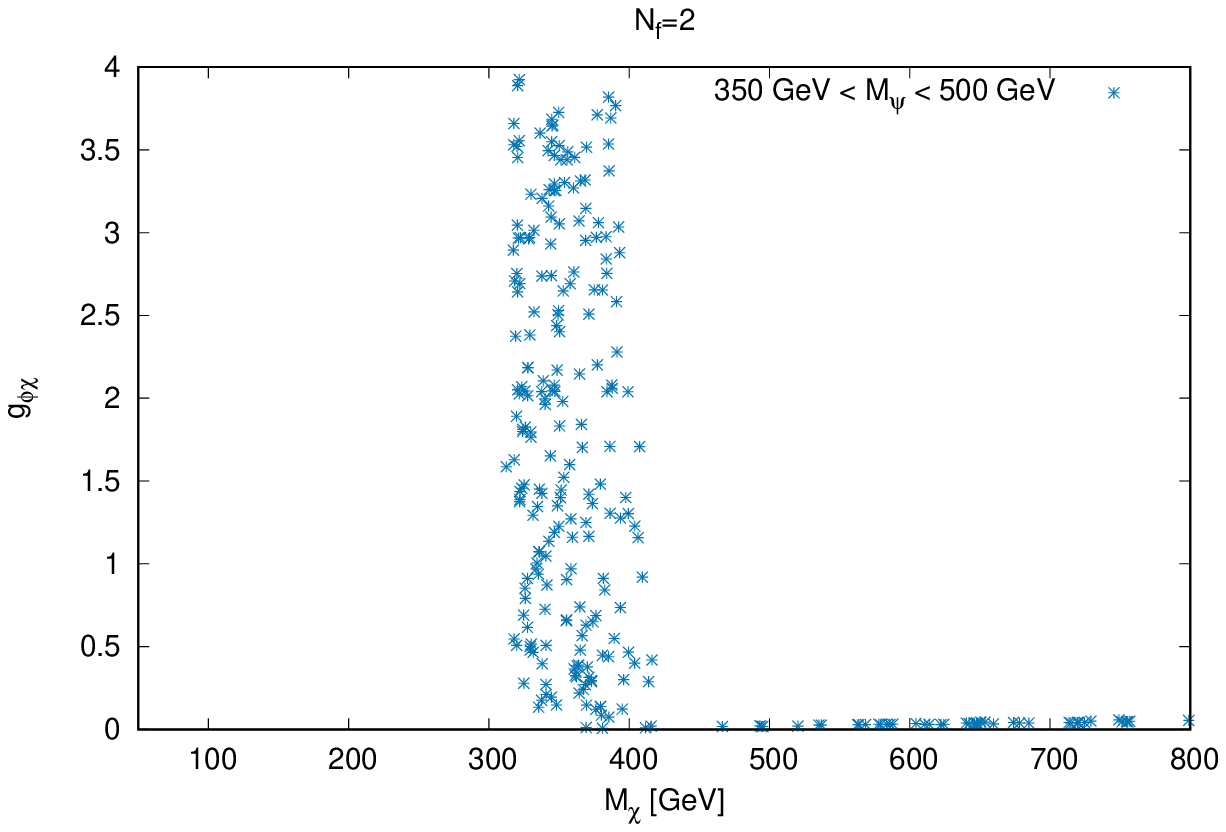,width=6.5cm}\hspace{0cm}\epsfig{figure=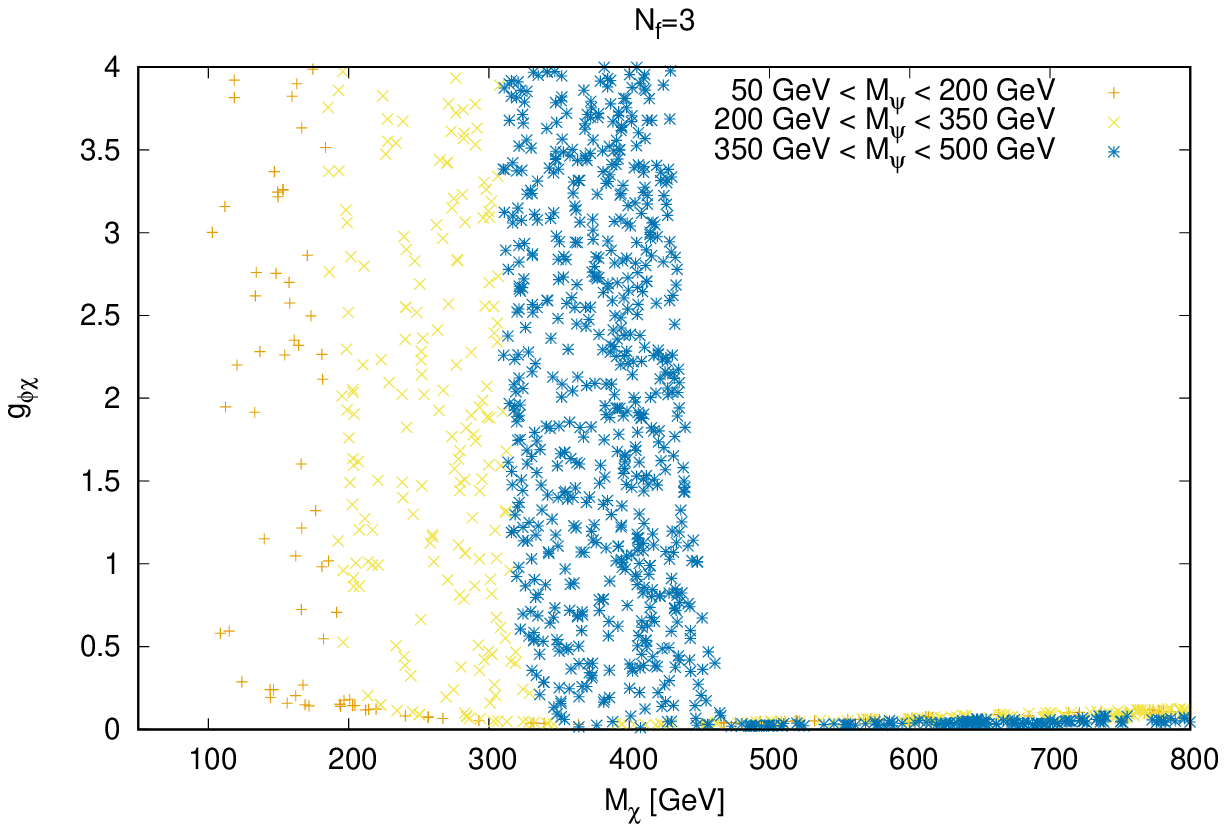,width=6.5cm}}
		\centerline{\vspace{-1.5cm}\hspace{0.5cm}(a)\hspace{4cm}(b)}
		\centerline{\vspace{0.5cm}}
	\end{center}
	\caption{Similar to Fig.\ref{scater-relic} except, we also consider the condition of the diphoton excess width at ATLAS.}\label{scaterrelicWidth}
\end{figure}

Studying a special case here, encourages us to imply the resonance condition on the relic density of DM to see the corresponding results. This analysis is depicted in Fig.~\ref{scaterrelicWidth} which makes the relevant parameter space (constructed in Fig.~\ref{scater-relic}) more accurate. Comparing with Fig.~\ref{scater-relic}, we see that increasing DM mass results in larger parameter space (same as the analysis in section~\ref{sec3}) but VLQs of first generation are suppressed here.

\section{Conclusion} \label{sec5}
We have proposed and analyzed a minimal model to explain new vector like fermions beyond the SM called VLQs. In this approach, we employed a singlet fermion as a DM candidate and following our aim, we defined new Dirac spinors playing the role of VLQs. The other new particle that plays a complementary role in the model is a pseudoscalar which mediates the interactions between dark and visible sectors. 

The first step was calculating the observed relic density of DM which is the most important constraint in DM phenomenology. Hence, we used the model independent free parameters and analyzed our results in the plane of DM mass and its coupling with the mediator ($M_\chi, |g_{\phi\chi}|$). This investigation also highlighted the role of the number of VLQs generations. 

In a resonace study, the mediator $\phi$ with mass of 750 GeV conducted us to obtain the desired cross section and decay width in the center of mass frame of colliding protons. Reaching this goal, we looked for an appropriate parameter space and formed it on the basis of VLQs mass and its coupling with the mediator ($M_\psi, |g_{\phi\psi}|$). A remarkable point in this probe was confirming the need for more generations of VLQs, so a better conscience with LHC bound was obtained.

In general, we expressed our results relying on two independent parameter-planes defined as ($M_\psi, |g_{\phi\psi}|$) and ($M_\chi, |g_{\phi\chi}|$) which are constructed from the model mass and coupling parameters. We also identified the number of VLQs generations needed for such an investigation. Finally, we concluded that such models are capable of producing the aforementioned quarks, have a good justification for a resonance at the LHC and can pass honorably the constraints coming from direct and indirect detections.

\end{document}